\documentclass[cite]{epl}

\usepackage{bm}

\usepackage{graphicx}

\title{Mechanism of Ground State Selection in the Frustrated Molecular Spin Cluster $\chem V_{15}$}

\shorttitle{Ground State Selection in $\chem V_{15}$}

\author{G.~Chaboussant\inst{1}\thanks{Present address: Laboratoire L\'eon Brillouin
(CNRS-CEA) CEA-Saclay, 91191 Gif-sur-Yvette Cedex, France. E-mail:
\email{chabouss@llb.saclay.cea.fr}}, S.T.~Ochsenbein\inst{1}, A.~Sieber\inst{1},
H.-U.~G\"{u}del\inst{1}, H. Mutka \inst{2}, A.~M\"{u}ller\inst{3} \and B. Barbara
\inst{4}}

\institute{ \inst{1} Department of Chemistry, University of Berne, CH-3000 Bern 9,
Switzerland. \\
\inst{2} Institut Laue-Langevin, BP 156, 38042 Grenoble, France. \\
\inst{3} Department of Chemistry, University of Bielefeld, 33501 Bielefeld, Germany. \\
\inst{4} Laboratoire de Magn\'etisme Louis N\'eel, CNRS, BP 166, 38042 Grenoble,
France. \\}

\shortauthor{G.~Chaboussant \etal}


\begin{document}

\maketitle

\begin{abstract}
We report an inelastic neutron scattering (INS) study under a magnetic field on the
frustrated molecular spin cluster $V_{15}$. Several field-dependent transitions are
observed and provide a comprehensive understanding of the low-energy quantum spin
states. The energy gap $2 \Delta_{0}\approx 27(3)\mu$eV between the two lowest
$S=1/2$ Kramers doublets is unambiguously attributed to a symmetry lowering of the
cluster. The INS data are mapped onto an S=1/2 Antiferromagnetic Heisenberg triangle
with scalene distortion. A quantitative description of the wavefunction mixing
within the ground state is derived.
\end{abstract}

Magnetic frustration operates when all the bonds in a magnetically coupled system
cannot be satisfied simultaneously. The ground state is then best described by a
superposition of quantum states with well defined probability of occurrence.
However, small irregularities like lattice strains, structural disorder or quantum
fluctuations can relieve part or all of the frustration, leading to a stabilised
ground state with a gap to less favourable ground state configurations ({\it order
by disorder} principle \cite{Villain}). The field of frustrated magnetism is
extremely active, encompassing low-dimensional materials \cite{Wessel01,Kodama02},
spin glasses or pyrochlores and Kagom\'e lattices \cite{review_frustration}. Most of
these materials are extended systems exhibiting short-range or quasi long-range
order, a situation rarely encountered in molecular magnets (or spin clusters) where
each such cluster in the lattice is magnetically well isolated from its neighbours
due to the presence of surrounding ligands. This magnetic shielding allows the study
of the individual behaviour of a finite number of interacting magnetic ions. Beside
being ideal candidates to study fundamental processes like quantum tunnelling or
quantum coherence at the nanoscale \cite{reviewSMM}, molecular magnetic clusters,
with their well defined nuclearities and topologies, also constitute quantum systems
in which geometrical frustration effects at the molecular level can be addressed.

A prominent example of a magnetically frustrated cluster is given by the the
polyoxovanadate complex $\chem K_{6}[V^{IV}_{15}As_{6}O_{42}(D_{2}O)] \cdot 8D_{2}O$
(hereafter $\chem V_{15}$) which contains 15 $V^{4+}$ spins ($S=1/2$) distributed
over two hexagons capping one triangle with global spherical shape
\cite{Muller88,Gatteschi91,Barra92}. The clusters have $D_{3}$ symmetry without
consideration of the hydrogen positions (inside and outside the cluster) and make up
a molecular crystal with trigonal symmetry. The $V^{4+}$ ions are
antiferromagnetically (AFM) coupled to their neighbours via oxo-bridges. Within the
hexagons, the AFM couplings are very strong (10-20 meV \cite{V15_exchange}) and the
spins on the triangle are coupled to the spins of the hexagons via frustrated
exchange couplings, but there are no significant {\it direct} exchange pathways
between the triangle spins. The coupling between the triangle spins occurs {\it
indirectly} through the hexagon spins. At low temperatures, the spins on the
hexagons are quenched in a singlet (S=0) state, and we are left with the three
triangle spins. The ground state is then made of two S=1/2 Kramers doublets
separated from the S=3/2 quartet state by about $0.315$ meV (3.7K)
\cite{Chiorescu,Chabouss02}. Inelastic neutron scattering (INS) has shown that the
two S=1/2 Kramers doublets are split by a gap $2\Delta_{0} \approx 35\mu$eV
\cite{Chabouss02}, and low-temperature magnetisation data were analysed in terms of
a two-level Landau-Zener model with a gap of about $7-8 \mu$eV ($80-100$ mK)
\cite{Chiorescu}. However, the microscopic origin of these gaps remained unclear.
Recent theoretical work suggested that Dzyaloshinskii-Moriya (DM) interactions might
be responsible for the gap opening
\cite{DMtheory_V15}. \\
In the present Letter, we report an INS study under a magnetic field of $\chem
V_{15}$ and show that deviations from trigonal symmetry are responsible for the
observed phenomena, not DM interactions. An analysis of both the energy levels and
the $Q$-dependence enables us to characterise the distortion and to address the
nature of the ground state.

The INS experiment was performed on the recently upgraded time-of-flight
spectrometer IN5 at the Institut Laue-Langevin (ILL, France) using cold neutrons of
wavelengths $\lambda = 7.5,9$ and $11${\AA}. Data were collected at temperatures
between $40$mK and $50$mK and corrected for the background and detector efficiency.
The magnetic field was provided by a $2.5$ Tesla superconducting coil. The
instrumental resolution at the elastic line ($\Gamma $, Full-Width at Half-Maximum)
is $\Gamma=28 \mu$eV at $7.5${\AA}, $\Gamma=18-21 \mu$eV at $9.0${\AA} and
$\Gamma=12 \mu$eV at $11${\AA}. We used a $4.6$ g polycrystalline powder sample of
fully deuterated $V_{15}$ placed under Helium in a rectangular flat Aluminum slab.

\begin{figure}
\onefigure[scale=0.85]{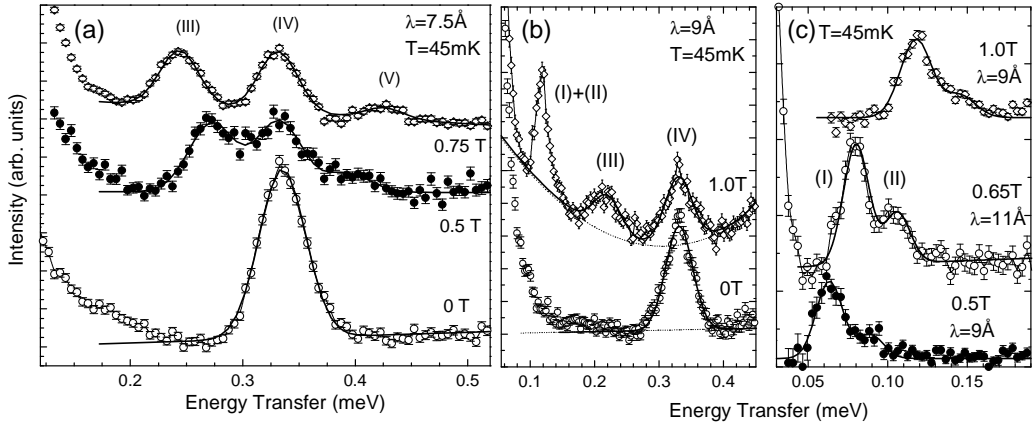}
\caption{(a) INS spectra at 0,0.5 and 0.75T ($\lambda = 7.5${\AA}). (b) INS spectra
at 0 and 1.0T ($\lambda = 9.0${\AA}) (c) INS spectra at 0.65T ($\lambda =
11.0${\AA}) and difference plots between non-zero magnetic field (H=0.5T, 1.0T) and
zero-field data obtained at $\lambda = 9.0${\AA}. Data in the $Q$-range between
0.2{\AA}$^{-1}$ and 1.1-1.4{\AA}$^{-1}$ were grouped together to improve statistics.
Solid lines are best fits to the data using Gaussian line shapes and a background.
Peaks are labeled as discussed in the text.} \label{fig1}
\end{figure}

\begin{figure}
\onefigure[scale=0.60]{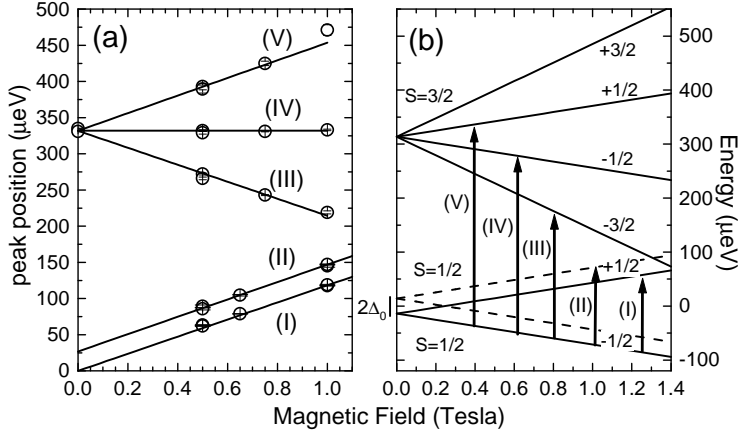}
\caption{(a) Field-dependence energies of the observed INS peaks (I) to (V). Solid
lines are linear fits to the data as discussed in the text. (b) Zeeman splittings
and assignment of the observed INS transitions.} \label{fig2}
\end{figure}

Figure~\ref{fig1} shows INS spectra obtained at $7.5${\AA} and $9.0${\AA} for
different values of the magnetic field. At H=0T, only one transition can be observed
at an energy of $\approx 0.335$ meV with $\Gamma \approx 41 \mu$eV. This width is
$1.5$ broader than the instrumental resolution and is intrinsic \cite{note1}. As the
field is switched on, satellite peaks appear {\it symmetrically} on each side of the
main peak but with different intensities. At $1$T, there is new intensity at about
$0.12$ meV. To better characterise it, the difference between the H=$0.5$/$1$T and
$0$T data is shown in fig.~\ref{fig1}c along with a higher resolution ($\lambda =
11${\AA}) spectrum obtained at $0.65$T. There are two peaks separated by about
$\approx 25-30 \mu$eV, and their energies increase linearly with magnetic field.
Their widths are found to be between $17$ and $19 \mu$eV, in the range of the
instrumental resolution, in contrast to the higher-energy peaks shown in
fig.~\ref{fig1}. Compiling the information derived from fig.~\ref{fig1}, it is
possible to draw the field-dependence of the INS transitions (see fig.~\ref{fig2}a)
where the five transitions are labeled (I) to (V).

To model our data, we consider the $S=1/2$ Heisenberg Antiferromagnetic (HAFM) model
on a triangle \cite{Gatteschi91,Barra92}:
\begin{equation}
{\cal H}_{0} = J_{12} \bm{S}_{1}\bm{S}_{2} + J_{23} \bm{S}_{2}\bm{S}_{3} + J_{13}
\bm{S}_{1}\bm{S}_{3} +  {\cal H}_{H} \; , \label{eq:Heis}
\end{equation}
where $\bm{S}_{1}$, $\bm{S}_{2}$, $\bm{S}_{3}$ denote the spin operators $1$,$2$ and $3$,
respectively and $J_{ij}$ is the Heisenberg exchange parameter between spins $i$ and $j$.
${\cal H}_{H} = \mu_{B}\bm{H} ( g_{1}\bm{S}_{1} + g_{2}\bm{S}_{2} + g_{3}\bm{S}_{3}) $
is the Zeeman interaction where the $g_{i}$-tensors are assumed to be diagonal. The
Zeeman splitting of the energy levels does not depend on the relative orientation to the
trigonal axis for isotropic couplings. For the equilateral HAFM triangle case ($J_{ij} =
J_{0} > 0$), the ground state consists of two degenerate S=1/2 Kramers doublets
separated from the $S=3/2$ excited state by $3J_{0}/2$. The wave functions of the two
degenerate S=1/2 Kramers doublets are given by $\Psi^{\pm \frac{1}{2}}_{0} = |
0,\frac{1}{2},\pm \frac{1}{2}>$ and $\Psi^{\pm \frac{1}{2}}_{1} = | 1,\frac{1}{2},\pm
\frac{1}{2}>$ in the basis $| S_{12},S,M >$ or any linear combination of them. If we
assume inequivalent couplings, there is a gap $2\Delta_{0}$ between the two Kramers
doublets but no splitting in the $S=3/2$ state. For instance, in the isosceles case
($J_{12}$ = $J$, $J_{13}$ = $J_{23}$ = $J'$) with $J > J'$, the energy gap becomes
$2\Delta_{0} = J-J'$ and $\Psi^{\pm \frac{1}{2}}_{0}$ is the lowest doublet.

From the field-dependence of the INS transitions shown in fig.~\ref{fig2}a, we can
immediately construct the energy diagram shown in fig.~\ref{fig2}b corresponding to
a distorted S=1/2 HAFM triangle with two S=1/2 doublet states and one S=3/2 quartet
state. Note that at $T=45$ mK, only the lowest doublet is populated in zero-field.
Transitions (I) and (II) are intra-doublet transitions with energies $\hbar
\omega_{I} = h = g\mu_{B}H$, $\hbar \omega_{II} = h + 2\Delta_{0}$, and transitions
(III) to (V) are transitions from the lowest doublet to the S=3/2 sublevels with
$\hbar \omega_{III} = 3J_{0}/2 + \Delta_{0} - h$, $\hbar \omega_{IV} = 3J_{0}/2 +
\Delta_{0} $ and $\hbar \omega_{V} = 3J_{0}/2 + \Delta_{0} + h$. The transition
$\hbar \omega = 2 \Delta_{0}$ between the two doublets at zero-field could not be
observed due to the width of the elastic line. The solid lines in fig.~\ref{fig2}a
corresponds to best fits with $g = 1.98(3)$, $2 \Delta_{0} = 27(3)\mu$eV $\approx
0.31(4)$K and $J_{0} = 212(2) \mu$eV $\approx 2.46$K. The g-factor value is in good
agreement with EPR \cite{Barra92,Ajiro01} and millimeter range spectroscopy
\cite{Vongtragool03}, the gap $2 \Delta_{0}$ is in relatively good agreement with
previous but less accurate INS measurements \cite{Chabouss02}, and the value of
$J_{0}$ is very close to the one inferred from magnetisation \cite{Chiorescu} and
INS \cite{Chabouss02}. So far, the isosceles triangle scenario successfully explains
the existing data, but there is no reason, {\it a priori}, to choose an isosceles
triangle: Lattice distortions, being static or dynamic, will most probably generate
a scalene triangle situation. From inspection of the energy levels it is not
possible to discriminate between isosceles and scalene triangles. We demonstrate now
that the INS intensities can discriminate between the two models.

The differential magnetic cross-section for a transition between initial state
$|S_{12},S,M>$ with energy $E_{s}$ and final state $|S'_{12},S',M'>$ with energy
$E_{f}$ is given by \cite{Lovesey}:
\begin{eqnarray}
& & \frac{d^{2}\sigma}{d \Omega dE}  = B_{\bm{Q}}\sum_{\alpha} \left( 1 -
\frac{Q^{2}_{\alpha}}{\bm{Q}^{2}} \right) \sum_{i,j}  e^{i \bm{Q} (\bm{R}_{i} -
\bm{R}_{j})}  \\ \nonumber & \times &  \langle S_{12},S,M |S^{\alpha}_{i}|
S'_{12},S',M' \rangle \langle S'_{12},S',M' |S^{\alpha}_{j}| S_{12},S,M \rangle
\times  \delta (\hbar\omega + E_{s} - E_{f}) \; , \\ \nonumber \label{INS_general}
\end{eqnarray}
with
\begin{equation}
B_{\bm{Q}} = \frac{N e^{- \beta E_{i}}}{Z} \frac{|\bm{k'}|}{|\bm{k}|} \cdot
F^{2}(\bm{Q}) \cdot e^{-2W(\bm{Q})} \; .
\end{equation}
$N$ is the number of magnetic centres in the sample, $\bm{k}$ and $\bm{k'}$ are the
initial and final neutron wave-vectors and $\bm{Q} = \bm{k} - \bm{k'}$ is the
scattering vector, $\exp[-2W(\bm{Q})]$ is the Debye-Waller factor, $F(\bm{Q})$ is
the magnetic form factor of the $\chem V^{4+}$ ions, $\bm{R}_{i}$ is the position of
the $i$th $\chem V^{4+}$ ion in the triangle, $\alpha = x,y$ or $z$, $\hbar\omega$
is the energy transfer and $Z$ the partition function. The matrix elements $\langle
S_{12},S,M |S^{\alpha}_{i}| S'_{12},S',M' \rangle$ are evaluated using irreducible
tensor operator (ITO) methods \cite{Budd63}. Only the triangular model is
considered. In the equilateral case, the {\it zero-field} and powder averaged
cross-section for the transition $\mid S_{12},S>$ to $\mid S'_{12},S'>$ is
\cite{Furrer79}:
\begin{equation} \frac{d^{2}\sigma}{d \Omega dE}  \sim  B_{\bm{Q}} \left( 1 -
\frac{\sin(QR)}{QR} \right) {\cal M}(S_{12},S,S'_{12},S') \; , \label{INS_equil_powder}
\end{equation}
where $R$ is the $V^{4+}$-$V^{4+}$ separation in the triangle and ${\cal M}$ is a
number that depends on the initial and final state quantum numbers. One can show
that it makes no difference to the INS intensities in zero magnetic field whether we
have $S_{12}=0$ or $S_{12}=1$ in the lowest doublet. In order to compare the
theoretical predictions with our experimental data, we need to consider the effect
of the magnetic field. This is done by using the Wigner-Eckart theorem
\cite{Furrer79}. This leads to INS intensities depending explicitly on $M,M'$ with
selection rules $M - M' = 0, \pm 1$. We obtain the following intensity ratios for
the doublet-quartet transitions (III:IV:V) = (3:2:1). The intra-doublet transition
with $\Delta S_{12}= S_{12}-S'_{12}=\pm 1$ has the same relative intensity as
transition (IV), whereas the $\Delta S_{12}=0$ transition is calculated to be three
times more intense. Comparison with the data in fig.~\ref{fig1} shows good, but not
perfect agreement for the doublet-quartet transitions, as (III) and (IV) have almost
equal intensity, instead of the calculated ratio $3/2$. From the relative intensity
of the intra-doublet transitions we can tentatively assign transition (I) to have
$\Delta S_{12}=0$ and transition (II) to have $\Delta S_{12}=\pm 1$. We already note
at this point, however, that transition (I) is weaker than the sum of (III), (IV)
and (V).

\begin{figure}
\onefigure[scale=0.55]{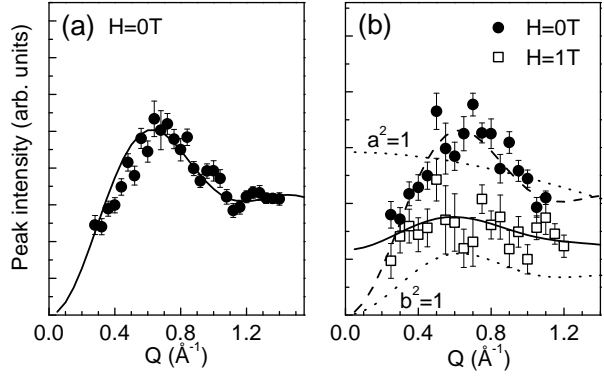}
\caption{(a) $Q$-dependence of the transitions III + IV + V measured at zero-field,
shown in fig.~\ref{fig1}. (b) $Q$-dependence of transition (I) at 1T and $\lambda =
9${\AA} (open squares) along with the III+IV+V transitions measured at 0T and at the
same wavelength. Lines (dashed and solid) are best fits using the calculated
$Q$-dependence (eq.~\ref{INS_equil_powder}) and the $V-V$ distance $R=6.92${\AA}.
Dotted lines correspond to theoretical curves with $a^{2}=1$ and $b^{2}=1$ (see
text). } \label{fig3}
\end{figure}

To get further insight into the nature of the transitions, in particular the
intra-doublet ones, we now consider their Q-dependence. Fig.~\ref{fig3}a shows the
$Q$-dependence of the sum of (III), (IV) and (V), measured at zero-field. The
agreement with the $Q$-dependence calculated with Eq.~\ref{INS_equil_powder} and the
$V$-$V$ distance $R=6.92${\AA} is very good, a confirmation that the S=3/2 state is
essentially unperturbed. Fig.~\ref{fig3}b shows the same transitions, {\it i.e.}
(III+IV+V) measured at zero-field and $\lambda = 9${\AA} (full circles), in
comparison with the $Q$-dependence of transition (I) at $1$T at the same wavelength
(open squares) \cite{note4}. While the doublet to quartet transition shows the same
$Q$-dependence as in fig.~\ref{fig3}a, the intensity of (I) is much less
$Q$-dependent, almost flat. Theoretically, an intra-doublet transition with $\Delta
S_{12}=\pm 1$ would be modulated by the $[1 - \sin(QR)/QR)]$ factor in
eq.~\ref{INS_equil_powder}. On the other hand, a $\Delta S_{12}=0$  transition does
not not have this factor and its $Q$-dependence is essentially flat. We have already
tentatively assigned band (I) to a Zeeman transition within the same doublet
($\Delta S_{12}=0$), and the $Q$-dependence now confirms this. However, we also
note, and fig.~\ref{fig3}b makes it very clear, that the overall intensity of peak
(I) is significantly smaller than the sum of (III+IV+V). The pure $\Delta S_{12}=0$
transition would have the same intensity, and the pure $\Delta S_{12}=\pm 1$ would
be three times weaker. This clearly suggests that the two doublets $\Psi^{\pm}_{0}$
and $\Psi^{\pm}_{1}$ are mixed in $\chem V_{15}$ as expected for a scalene triangle.
Defining the lower-lying doublet as $\Omega^{\pm}_{0} = a \Psi^{\pm}_{0} + b
\Psi^{\pm}_{1}$ (with $a^{2} + b^{2} = 1$), we get the following $Q$-dependent
intensity for transition (I):
\begin{equation}
I_{I}(Q)  =  I_{0} F^{2}(\bm{Q}) \left[ a^{2} +  \frac{b^{2}}{3} \left( 1 -
\frac{\sin(QR)}{QR} \right) \right] \; ,
\end{equation}
where $I_{0}$ is an intensity factor, kept fixed to the value determined from the
intensity of (III+IV+V) at the same wavelength. The dotted lines in fig.~\ref{fig3}b
represent the disentangled situations ($[a,b]$=$[1,0]$ or $[0,1]$)  and the full
line is a best fit to the data corresponding to $a^{2}=0.4$ and $b^{2}=0.6$. The
mixing of the $S_{12}=0$ and $S_{12}=1$ states is quite substantial. One parameter
set that produces this situation is $J_{12} = 0.21$ meV, $J_{23} = 0.23$ meV and
$J_{13} = 0.20$ meV. This set is not unique as we only have access to $a^2$ and
$b^2$, but it gives a clear idea of the exchange coupling variation caused by the
distortion.

We now briefly discuss the relevance of Dzyaloshinskii-Moriya (DM) interactions
\cite{DMtheory_V15}. The effect of DM interactions in the S=1/2 AFM triangle has
been treated in some detail \cite{DM-Tsukerblat} and we give here the main results.
The DM interaction introduces off-diagonal matrix elements which lead to a first
order splitting $2 \Delta$ of the doublets. If we assume perfect triangular
symmetry, we have $2 \Delta = d_{z}/ \sqrt{3}$ \cite{Boca} where $d_{z}$ is the DM
parameter. If the splitting in $\chem V_{15}$ was due to DM interactions this would
lead to $d_{z} = 47 \mu$eV. In contrast to the pure Heisenberg case, the energy
levels depend explicitly on the relative orientation of the clusters to the magnetic
field in this case \cite{DM-Tsukerblat}. In a powder measurement, the overall
spectrum will reflect this field dependence by a broadening of all the peaks, in
particular the intra-doublet transitions. This is clearly not what we observe
experimentally. In addition, the theoretical energy difference $2 \Delta$ in the DM
model between transitions (I) and (II) is field dependent, decreasing from $27
\mu$eV at $0$T to $17 \mu$eV at 1T for a $d_{z}$ value of $47\mu$eV. Again, this is
completely incompatible with our experimental results, which show a field
independent energy difference between (I) and (II) as well as a linear field
dependence of transitions (III),(IV) and (V), see fig.~\ref{fig2}. DM interactions
can thus definitely be ruled out as the origin of the ground state splitting in
$\chem V_{15}$.

The present study demonstrates that the low-energy properties of $\chem V_{15}$ are
accurately described by a triangle model with {\it scalene} distortion. In
particular, the mechanism generating a gap $2 \Delta_{0} = 27(3) \mu$eV between the
two Kramers doublets is unambiguously established using both energy and wavefunction
information provided by INS. The symmetry lowering is already established from the
field dependence of the energy levels. However, {\it only} inspection of the
intensities and their $Q$-dependence leads to a detailed knowledge of the
wavefunction mixing within the ground state resulting from the scalene distortion of
the triangle. An obvious source of symmetry lowering is provided by the water
molecule located in the center of the spherical cavity of $\chem V_{15}$. Another
one is given by some disorder (partial occupancy) on the water structure of the
lattice \cite{Gatteschi91,Barra92,Choi03}. This study shows that the impact of very
small structural perturbations to relieve magnetic frustration can be sizeable.
$\chem V_{15}$ constitutes a example of a nanometer-scale system with maximised
magnetic frustration (triangle) where {\it order}, {\it i.e.} a stabilised ground
state with a gap to less favourable "ground states", is induced by a small
structural distortion ({\it disorder}).

\begin{acknowledgments}

We are grateful to F. Mila and M. Elhajal for stimulating discussions, to J.-L.
Ragazzoni and S. Jenkins for their technical support at the ILL and to E.
Krickemeyer and H. B\"ogge for help in the synthesis. We wish to thank J. Ollivier,
M. Plazanet and the IN5 team at the ILL for the very significant improvement of the
spectrometer. One of us (A.M) thanks the Deutsche Forschungshemeinschaft and the
Fonds der Chemischen Industrie for financial support. This work has been supported
by the Swiss National Science Foundation and by the TMR program Molnanomag of the
European Union (No: HPRN-CT-1999-00012).
\end{acknowledgments}


\begin{thebibliography}{}

\bibitem{Villain} \Name{Villain J.} \REVIEW{Z. Physik B}{33}{1979}{31};
\Name{Villain J., Bideaux R., Carton J.P \and Conte R.} \REVIEW{J. Phys. (Paris)}
{41}{1980}{1263}.

\bibitem{Kodama02} \Name{Kodama K., Takigawa M., Horvatic M., Berthier C.,
Kageyama H., Ueda Y., Miyahara S., Becca F. \and Mila F.} \REVIEW{Science}
{298}{2002}{395} and references therein.

\bibitem{Wessel01} \Name{Wessel S., Normand B., Sigrist M. \and Haas S.}
 \REVIEW{Phys. Rev. Lett.}{86}{2001}{1086}.

\bibitem{review_frustration} For a review see \Name{Ramirez A.P} in
 \REVIEW{{\it Handbook on Magnetic Materials}}{13}{1999}{423} (Elsevier Science,
Amsterdam); \Name{Bramwell S.T. \and Gingras M.J.P.}
\REVIEW{Science}{294}{2001}{1495}; \Name{Moessner R.} \REVIEW{Can. J.
Phys.}{79}{2001}{1283}.

\bibitem{reviewSMM} \Name{Barbara B., Thomas L., Lionti F., Chiorescu I. \and Sulpice A.}
 \REVIEW{J. Mag. Magn. Mat}{200}{1999}{167}; \Name{Sessoli R. \and Gatteschi D.}
 \REVIEW{Angew. Chem. Int. Edit.}{42}{2003}{268}.

\bibitem{Muller88} \Name{M\"{u}ller A. \and D\"{o}ring J.}
 \REVIEW{Angew. Chem., Int. Ed. Engl.}{27}{1988}{1721}.

\bibitem{Barra92} \Name{Barra A. L., Gatteschi D., Pardi L., M\"{u}ller A. \and D\"{o}ring J.}
 \REVIEW{J. Am. Chem. Soc.}{114}{1992}{8509}.

\bibitem{Gatteschi91} \Name{Gatteschi D., Pardi L., Barra A. L. , M\"{u}ller A. \and D\"{o}ring J.}
 \REVIEW{Nature}{354}{1991}{463}.

\bibitem{V15_exchange} \Name{Kostiyuchenko V.V \and Zvezdin A.K.}
 \REVIEW{Phys. Sol. State}{45}{2003}{903};

\Name{Platonov V.V., Tatsenko O.M., Plis V.I., Popov A.I., Zvezdin A.K. \and Barbara
B.} \REVIEW{Phys. Stat. Sol.}{44}{2002}{2104}.

\bibitem{Chabouss02} \Name{Chaboussant G., Basler R., Sieber A., Ochsenbein S.T, Desmedt A.,
Lechner R.E., Telling M. T. F., K\"{o}gerler P., M\"{u}ller A. \and G\"{u}del H.-U.}
\REVIEW{Europhys. Lett.}{59}{2002}{291}.

\bibitem{Chiorescu} \Name{Chiorescu I., Wernsdorfer W., M\"{u}ller A., B\"{o}gge H. \and Barbara B.}
 \REVIEW{J. Mag. Mag. Mat.}{221}{2000}{103}; \Name{Chiorescu I., Wernsdorfer W.,
M\"{u}ller A., B\"{o}gge H. \and Barbara B.} \REVIEW{Phys. Rev.
Lett.}{85}{2000}{3454}; \Name{Chiorescu I., Wernsdorfer W., M\"{u}ller A., Miyashita
S. \and Barbara B.} \REVIEW{Phys. Rev. B}{67}{2003}{020402}.

\bibitem{DMtheory_V15} \Name{Konstantinidis N.P. \and Coffey D.}
 \REVIEW{Phys. Rev. B}{66}{2002}{174426}; \Name{De Raedt H., Miyashita S.
 \and Michielsen K.} cond-mat 0306275.

\bibitem{Ajiro01} \Name{Ajiro Y. Itoh H., Inagaki Y., Asano T., Narumi Y., Kindo K.,
Sakon T., Motokawa M., Cornia A., Gatteschi D., M\"{u}ller A. \and Barbara B.} in
Proceedings of French-Japanese Symposium at Fukuoka , Japan (2001).

\bibitem{note1} At $\lambda = 9.0${\AA}, the resolution is better and, at zero-field,
the width of the main line is still around $40 \mu$eV. Therefore the intrinsic width of
the $0.335$ meV peak is $40 \mu$eV.

\bibitem{Vongtragool03} \Name{Vongtragool S., Gorshunov B., Mukhin A.A., Van Slageren J.,
Dressel M. \and  M\"{u}ller A.} \REVIEW{Phys. Chem. Chem. Phys.}{5}{2003}{2778}.

\bibitem{Lovesey} \Name{Lovesey S.M.} Theory of Thermal Neutron Scattering from Condensed Matter,
(Clarendon Press Oxford), 1984.

\bibitem{Budd63} \Name{Budd B.R.} Operator Techniques in Atomic Spectroscopy (McGraw-Hill, New-York,
1963).

\bibitem{Furrer79} \Name{Furrer A.\and G\"{u}del H.-U.} \REVIEW{Phys. Rev. Lett.}{39}{1977}{657};
\Name{Furrer A. \and G\"{u}del H.-U.} \REVIEW{J. Mag. Mag. Mat.}{14}{1979}{256}.

\bibitem{note4} At $\lambda = 7.5${\AA} transition (I) is in the tail of the elastic line
and it is not possible to obtain a reliable $Q$-dependence of this transition with
this setting. Thus, to compare its Q-dependence with the main transition at
zero-field, we consider the $\lambda = 9.0${\AA} data where the two transitions are
observed under the same conditions.

\bibitem{DM-Tsukerblat} \Name{Tsukerblat B.S., Kuyavskaya B.Y., Belinskii M.I.,
Ablov A.V., Novotortsev V.M. \and Kalinnikov V.T.} \REVIEW{Theoret. Chim. Acta
(Berl.)} {38}{1975}{131}; \Name{Rakitin Y.V., Yablokov Y.V. \and Zelentsov V.V.}
\REVIEW{J. Magn. Res.}{43}{1981}{288}.

\bibitem{Boca} \Name{Bo\v{c}a R.} Theoretical Foundations of Molecular Magnetism,
Elsevier (1999).

\bibitem{Choi03} \Name{Choi J., Sanderson L.A.W., Musfeldt J.L., Ellern A.,
\and K\"{o}gerler P.} \REVIEW{Phys. Rev. B.}{68}{2003}{064412}.

\end{thebibliography}
\end {document}